\documentclass[useAMS,usenatbib]{mnras}  
\usepackage{amssymb}
\usepackage{amsfonts}
\usepackage{tabularx}
\usepackage{graphicx}
\usepackage{ulem}
\usepackage{array}
\usepackage{color}
\usepackage{booktabs}
\usepackage[fleqn]{amsmath}
\usepackage{subfigure}

\usepackage{mathtools}

\makeatletter
\newcommand*{\rom}[1]{\expandafter\@slowromancap\romannumeral #1@}
\makeatother
\numberwithin{equation}{section}
\def\d{{\rm d}}

\newcommand{\p}{\partial}

\newcommand{\bv}{{\mbox{\boldmath $v$}}}

\newcommand{\br}{{\mbox{\boldmath $r$}}}
\newcommand{\ba}{{\mbox{\boldmath $a$}}}

\renewcommand{\sovast}{Soviet Astronomy}

\begin{document}

\title[Dominating harmonic potential]{Simulation of the loss-cone instability in spherical systems. I. Dominating harmonic potential}
\author[E.\,V.\,Polyachenko et al.]
   { E.~V.~Polyachenko\,$^1$\thanks{E-mail: epolyach@inasan.ru},
     P.~Berczik\,$^{2,3,4}$\thanks{E-mail: berczik@mao.kiev.ua},
     A.~Just\,$^3$\thanks{E-mail: just@ari.uni-heidelberg.de},
     I.~G.~Shukhman\,$^5$\thanks{E-mail: shukhman@iszf.irk.ru}\\
     $^1 ${\it\small Institute of Astronomy, Russian Academy of Sciences,}
          {\it\small 48 Pyatnitskya St., Moscow 119017, Russia}\\
     $^2 ${\it\small The International Center of Future Science of the Jilin University, 2699 Qianjin St., 130012 Changchun City, PR China }\\
     $^3 ${\it\small Zentrum f\"ur Astronomie der Universit\"at Heidelberg, Astronomisches Rechen-Institut,  }
          {\it\small M\"{o}nchhofstr. 12-14, 69120 Heidelberg, Germany} \\
     $^4 ${\it\small Main Astronomical Observatory, National Academy of Sciences of Ukraine, MAO/NASU,}
          {\it\small 27 Akad. Zabolotnoho St. 03680 Kyiv, Ukraine} \\
     $^5 ${\it\small Institute of Solar-Terrestrial Physics, Russian Academy of Sciences,}
          {\it\small Siberian Branch, P.O. Box 291, Irkutsk 664033, Russia }}



\maketitle

\begin{abstract}
A new so-called `gravitational loss-cone instability' in stellar systems has recently been investigated theoretically in the framework of linear perturbation theory and proved to be potentially important in understanding the physical processes in centres of galaxies, star clusters, and the Oort comet cloud. Using N-body simulations of a toy model, we confirm previous findings for the harmonic dominating potential and go beyond the linear theory. Unlike the well-known instabilities, the new one shows no notable change in the spherical geometry of the cluster, but it significantly accelerates the speed of diffusion of particles in phase space leading to an early instability saturation.
\end{abstract}


\section{Introduction}
\label{sec:intro}%

Stellar systems, in general, are rich in different kinds of instabilities~\citep{1984sv...bookQ....F} that are the fastest dynamical change driving processes, leaving behind different relaxation mechanisms~\citep{2005ApJ...625..143T}. In particular, spheres are often subject to the well-known radial-orbit instability (ROI) \citep{PS1972, A73, PS1981SvA, 1994ASSL..185.....P} inherent in radially anisotropic systems \citep[for a review, see][]{2015MNRAS.451..601P}.

When ROI is out of play, another mechanism called gravitational loss-cone instability (gLCI) can be important. The first example was given in \citet{1991SvAL...17..371P} by investigating a simple analytical disc model. This and other studies \citep{2005ApJ...625..143T, 2007MNRAS.379..573P} assume that a cluster of mass $M_*$ is embedded in the dominating potential of a central point mass $M$, so that $\varepsilon\equiv M_*/M$ is a small parameter. It allows us to consider the slow precessing motion of stellar orbits and to go far enough in analytics. The typical times for precession, $t_{\rm pr}$, and instability, $t_{\rm ins}$, are:
\begin{equation}
    t_{\rm pr} \sim t_{\rm ins} \sim \varepsilon^{-1} t_{\rm dyn}\,,
\end{equation}
i.e. large in units of the dynamical time $t_{\rm dyn}\sim (GM/R^3)^{-1/2}$, where $R$ is the cluster radius.

Another dominating potential leading to slow orbital precession is produced by an extended homogeneous halo, which is harmonic, $\Phi_0 = \Omega_0^2 r^2/2$, with $\Omega_0$ being the orbital frequency, or inverse dynamical time. Substitution $M \to R^3\Omega_0^2/G$ establishes a connection between these two cases. It is worthwhile to note that the harmonic potential is not only for academic purposes: observations also suggest that central dark matter density of galaxies most probably have constant density cores on the scale of $\sim 1$\,kpc~\citep{2006MNRAS.373.1451R}.

The instability strength is characterized by the ratio of the exponential growth rate $\gamma$ to the dynamical frequency, or e-folding time to dynamical time. The typical values of the latter for bar instability is $2\,-\,10$, for ROI $\gtrsim 1$, for the gLCI $(5\,-\,10)\,\varepsilon^{-1}$. The time interval of numerical simulations to catch the gLCI thus should be by a factor of $\varepsilon^{-1}$ longer than ordinary simulations of galactic discs. Furthermore, depending on the orbit integration scheme in particular N-body realizations, the numerical workload can be extremely high, significantly constraining the allowed number of particles.

As a first step to simulate the gLCI, we adopt a toy model in a dominating harmonic potential, as described in \citet{2010AstL...36..175P}. The model details and N-body set-up are given in Section 2. Results containing a comparison of the instability growth rate with the linear perturbation theory, and analysis of peculiarities of the cluster evolution are given in Section 3. The final section 4 contains conclusions, final remarks, and describes future perspectives.

\section{The model and set-up}
\label{sec:nbody-set}

\citet{2010AstL...36..175P} considered the one-parameter family of distribution functions (DF) with power law dependence on angular momentum
\begin{equation}
    F(E,L) = N_n \delta(E)\,\alpha^n\,,\quad\alpha\equiv L/L_{\rm circ}\,,
    \label{eq:df}
\end{equation}
where $E=v^2/2+\Phi(r)$ and $L=|\br\times\bv|$ are (specific) energy and angular momentum of a particle; $N_n$ is the normalization constant; $L_{\rm circ}$ is the angular momentum of the circular orbit, corresponding to $E=0$. The Dirac delta-function $\delta(x)$ implies that all particles have the same energy $E=0$ (mono-energetic). The potential
\begin{equation}
    \Phi(r) = \frac{\Omega_0^2 r^2}2 + \Phi_*(r) + {\rm constant}\,,
\end{equation}
where the constant in the r.h.s. is chosen so that $\Phi(r)$ vanishes on the edge of the sphere $R$; $\Phi_*(r)$ is the potential due to the cluster, which density distribution is:
\begin{equation}
    \rho(r) \propto r^n (R^2-r^2)^{(n+1)/2}\,.
    \label{eq:rho}
\end{equation}

In the models with slow precessional motion of orbits, the perturbed DF allows averaging over the fast dynamical period. In this case, the dynamics of the system is mainly determined by the DF dependence on the angular momentum, while the energy dependence of the unperturbed DF
is not significant~\citep{2005ApJ...625..143T, 2007MNRAS.379..573P}. On the other hand, the mono-energetic DF is preferable for analytical studies. The power law loss cone $\propto L^n$ can be astrophysically justified only for nearly radial orbits. In our previous theoretical study and here we expanded the law up to the circular orbits for simplicity. Note  that  precession can also be due to a weak radial dark matter halo inhomogeneity, for any light embedded stellar cluster \citep{2010AstL...36..175P}.

The DFs (\ref{eq:df}) belong to the family of generalised power laws, for which the local anisotropy parameter $\beta(r) \equiv 1-\sigma_p^2/\sigma_r^2 = -n/2$.  According to the theoretical criterion, gLCI is possible when the orbit precession is retrograde, providing the DF is a growing function of angular momentum. Orbits with retrograde precession occur for $n>0.4$, i.e. all models in the family interesting for our simulations are tangentially anisotropic.

\begin{figure}
\centering
  \centerline{\includegraphics[width=\linewidth]{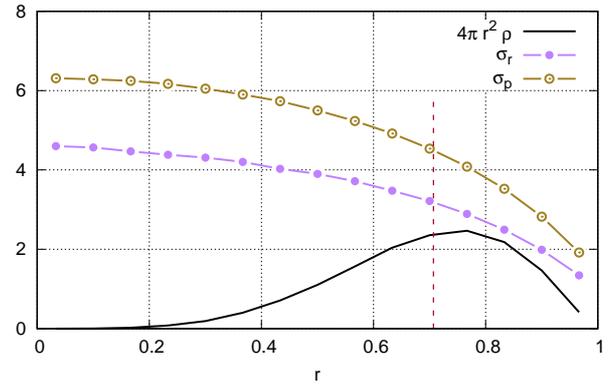}}
  \caption{The numerical model: initial mass distribution and velocity dispersion profiles for radial and transversal directions. The vertical dashed line shows the circular orbit radius  $r_{\rm circ}$.}
  \label{fig:model}
\end{figure}

\begin{figure}
\centering
  \centerline{\includegraphics[width=\linewidth]{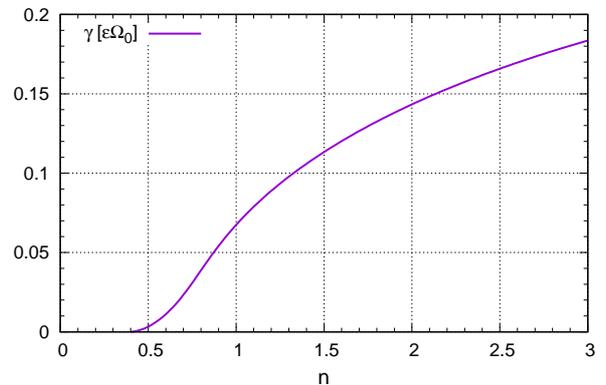}}
  \caption{The exponential growth rate vs. index $n$ for models of Eq.\,\ref{eq:df} in instability time units.}
  \label{fig:gr}
\end{figure}

In what follows, we set $G=M_*=R=1$, $\Omega_0=10$, so that the small parameter $\varepsilon=0.01$. We also fix a typical value for the index $n=2$. The initial density and velocity dispersion profiles $\sigma_r$ and $\sigma_p$ ($=\sigma_\theta=\sigma_\varphi$) are shown in Fig.\,\ref{fig:model}. The linear theory predicts an exponential growth rate for the quadrupole harmonic of perturbations $\gamma \approx 0.14\,\varepsilon\,\Omega_0$ (see Fig.\,\ref{fig:gr}), the corresponding e-folding time is $\approx 70$.

After several auxiliary runs aimed to determine the optimal code and integration scheme, time steps and gravity softening, we choose the leap-frog integrator and gravity softening equal to 0.001. The simulations were performed using the own developed TREE-GPU code {\tt ber-gal} \citep{2015ApJ...806..267Z, PBJ16, 2019A&A...622L...6K} with an opening angle $\theta$ = 0.5. The number of particles varied in the range from $0.25 \times 10^6$ to $4 \times 10^6$. The current set of simulations was carried out with the new GPU version of the code using the very recent NVIDIA Graphics Processing Unit (GPU) platform. On a typical desktop hardware (CPU: i5-2500K with 4 cores @ 3.3 GHz + GPU: GeForce GTX 570 with 480 cores @ 1.46 GHz) we get the results for the full self-gravity force calculation routine 
for $N= 4\times 10^6$ particles in $\approx$ 8 sec. The typical $N= 4\times 10^6$ model presented in the current paper is run up to $t = 500$ time units in $\sim$ 68 hours.

The initial N-body realisation was obtained using the standard von Neumann rejection technique in $L$-space, which fixes the orbital shape. The orbit was oriented randomly in spherical coordinates, while the position along the orbit was chosen inversely proportional to the radial velocity. Due to numerical noise, a position of the centre of mass was not exactly zero. In some runs, we performed a centre-of-mass correction,
however with no impact on the results.

\section{Results}
\label{sec:nbody-res}

The stability analysis considers spherical harmonics of perturbations separately. The density distribution $\rho(r,\theta,\varphi)$ can be represented as follows
\begin{equation}
\rho(r,\theta,\varphi)=\sum\limits_{l=0}^{\infty} \rho_l(r,\theta,\varphi)\,,
\label{eq:rho}
\end{equation}
where
\begin{equation}
\rho_l(r,\theta,\varphi)=\sum\limits_{m=-l}^l C_l^m(r)\, Y_l^m(\theta,\phi)\,,
\end{equation}
and $Y_l^m(\theta,\phi)$ are spherical harmonics. The expansion coefficients $C_l^m$ integrated over radius give a global characteristic amplitude $A_l$ of spherical harmonic $l$, see details in Appendix \ref{app1}.

Fig.\,\ref{fig:a2} presents the growth of the quadrupole harmonic amplitude $A_2$ extracted from the N-body simulations. This harmonic, as well as other spherical harmonics, has a non-zero initial value due to numerical noise. After some adjustment, it begins to grow exponentially in good agreement with our previous theoretical findings. However, after $t=350$ the instability is attenuated and then saturates barely reaching the level $0.02$.

\begin{figure}
\centering
  \centerline{\includegraphics[width=\linewidth]{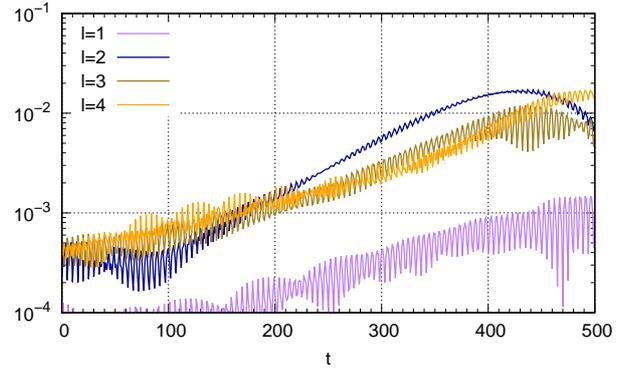}}
  \caption{Growth of the amplitudes of the lowest spherical harmonics ($l\le 4$, $N=4\times 10^6$). The low initial amplitude of the dipole perturbation ($l=1$) is due to centre-of-mass correction of the initial data.}
  \label{fig:a14}
\end{figure}

The lowest order spherical harmonics show growth rates and saturation levels similar to the quadrupole term (Fig.\,\ref{fig:a14}). The instability has no strong effect on the shape of the cluster, which 
remains almost spherical (but for the radial structure see Fig.\,\ref{fig:rho_t} below).

\begin{figure}
\centering
  \centerline{\includegraphics[width=\linewidth]{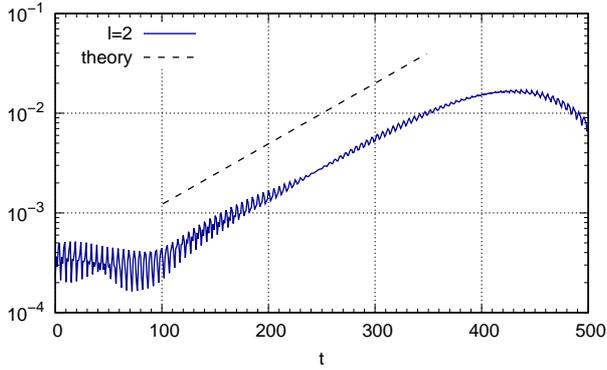}}
  \caption{Growth of the amplitude of the quadrupole harmonic ($N=4\times 10^6$). The theoretically predicted growth rate corresponds to the slope shown by the dashed line.}
  \label{fig:a2}
\end{figure}

\begin{figure}
\centering
  \centerline{\includegraphics[width=\linewidth]{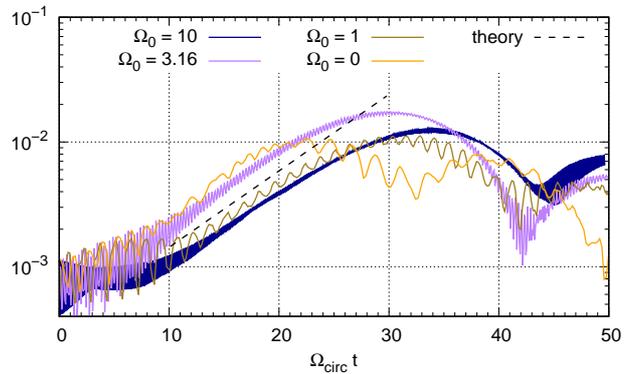}}
  \caption{Amplitudes of the quadrupole harmonics for different external halo parameters $\Omega_0$, as function of normalised time units ($N=0.25\times 10^6$).}
  \label{fig:A2k}
\end{figure}

{ A variation of the $\Omega_0$ parameter shows that the growth rate of instability does not depend strongly on the strength of the external homogeneous halo. Fig.\,\ref{fig:A2k} compares the slopes for several runs in units of dynamical times, which is here characterized by $\Omega_{\rm circ} = L_{\rm circ}/r_{\rm circ}^2$ rather than $\Omega_0$ since it suits better for weak external halos (but close to $\Omega_0$ for heavy halos). The growth rates and attenuation of the instability occur similarly.}

\begin{figure*}
\centering
  \centerline{\includegraphics[width=\linewidth]{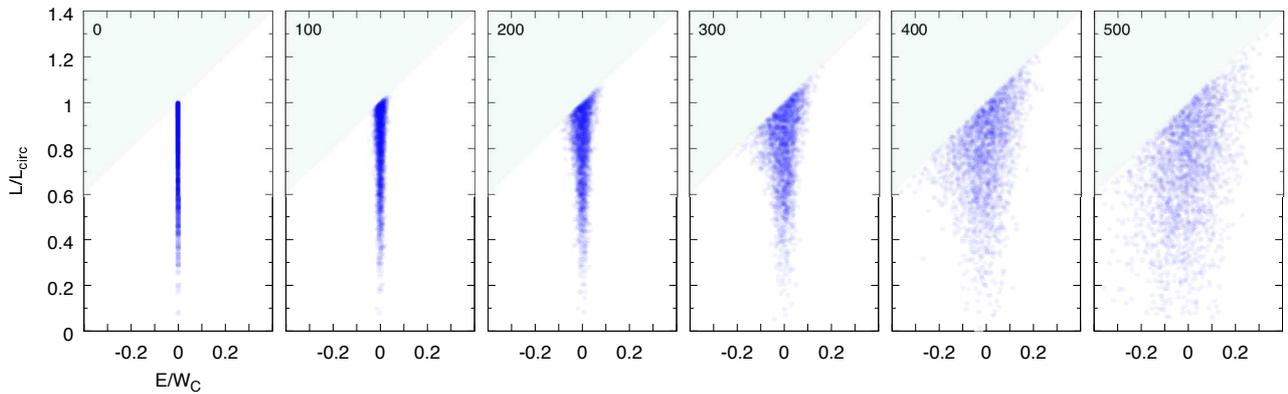}}
  \caption{Spreading of the initial DF in $E/W_{\rm C}-L/L_{\rm circ}$ phase space in N-body simulations ($N=4\times 10^6$). The shadowed area is above the circular orbit's curve, i.e. unavailable for particles. The simulation time of the snapshot is shown in the upper left. Here $W_{\rm C}$ is the virial of Clausius, $W_{\rm C} = \left|\sum_i \br_i \times \ba_i \right|/N$, where $\ba_i$ is the acceleration of the particle $i$ ~\citep[e.g.,][]{2015MNRAS.453.2919S}.}
  \label{fig:smearing}
\end{figure*}

To find a reason for such an early instability saturation, we analysed the particle distribution in phase space. The initial DF represents a narrow line in the $(E,L)$-plane (Fig.\,\ref{fig:smearing}, left panel). In absence of collisions and instability, the DF should remain fixed over time, since $E$ and $L$ are constants of motion. Naturally, due to various diffusion processes, a spreading of the narrow distribution occurs. However, the observed relaxation has been too fast. Indeed, the two-body relaxation time for a Maxwellian DF is \citep[e.g.,][p.587]{BT08}:
\begin{equation}
    t_{\rm rel} = 0.34 \frac{N\sigma_0^3}{G^2 M_* \rho \ln\Lambda}\,,
\end{equation}
where $\ln \Lambda $ is the Coulomb logarithm, $\sigma_0$ is the velocity dispersion. For the shortest possible time estimate, assume $\sigma_0=\sigma_r$, and $\ln \Lambda = \ln N$. Then for $N=4\times 10^6$ we estimate $t_{\rm rel} \sim 10^7$, which is by far larger than the time of the simulation.

\begin{figure}
\centering
  \centerline{\includegraphics[width=\linewidth]{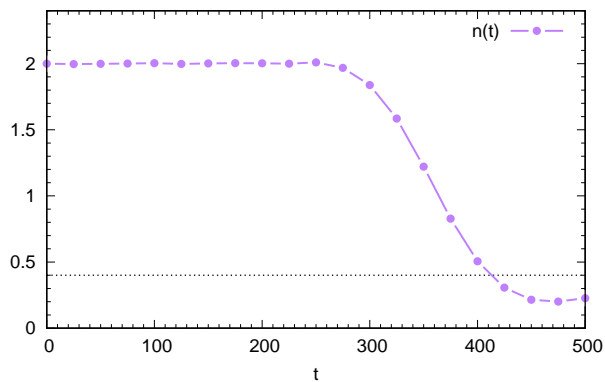}}
  \caption{Fitted effective index $n$ of the model (Eq.\,\ref{eq:df}) vs. time.}
  \label{fig:nt}
\end{figure}

Fig.\,\ref{fig:rho_t} presents the radial density profile change during the simulation. Up to $t=300$, the profile doesn't change; after that, a homogeneous core begins to form. { Meanwhile, the DF changes towards more isotropic models: fitting the angular momentum distribution in the range $0<\alpha<3/4$ shows a drop of the index $n$ from 2 down to 0.2, meaning attenuation and cease of the instability (Fig.\,\ref{fig:nt})}.
\begin{figure}
\centering
  \centerline{\includegraphics[width=\linewidth]{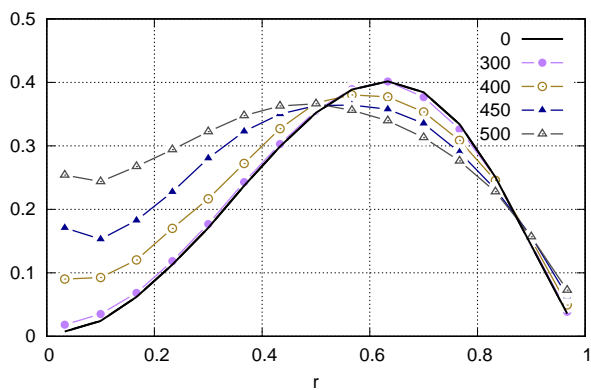}}
  \caption{Time dependence of the density $\rho(r)$ for selected N-body snapshot times ($N=4\times 10^6$).}
  \label{fig:rho_t}
\end{figure}

In systems with dominating Keplerian or harmonic potential, a more efficient mechanism called resonant relaxation should take place~\citep{1996JRASC..90..334R}. This relaxation does not affect spreading in energy but shortens the relaxation time in angular momentum by a factor $\varepsilon \ln\Lambda$ (=0.15 for $N=4\times 10^6$).

To quantify the speed of the diffusion in $E$ and $L$, we evaluated the spread $\sigma_t[E]$ as the difference between the third and the first quartiles of the distribution of energy of individual particles $E_i(t)$. This method is more robust to outliers than the calculation of standard deviation $\sigma$, while it gives $1.35\,\sigma$ for the normal distribution. Similarly, $\sigma_t[L]$ was obtained for the angular momentum shifts $L_i(t)-L_i(0)$. In the diffusion driven by two-body relaxation, we expect
\begin{equation}
    \frac{\sigma^2_t[E]}{W^2} = \zeta \frac{t}{t_{\rm rel}}\,,\quad \frac{\sigma^2_t[L]}{L^2_{\rm circ}} = \eta \frac{t}{t_{\rm rel}}\,
\end{equation}
where $\zeta$, $\eta$ are constants of order unity, $W$ is some energy characteristic of the system, e.g. total kinetic energy, or virial of Clausius $W_{\rm C}$~\citep[e.g.,][]{2015MNRAS.453.2919S}. If the resonant relaxation takes place, we expect $\eta \sim (\varepsilon \ln\Lambda)^{-1}$.

\begin{figure}
\centering
  \centerline{\includegraphics[width=\linewidth]{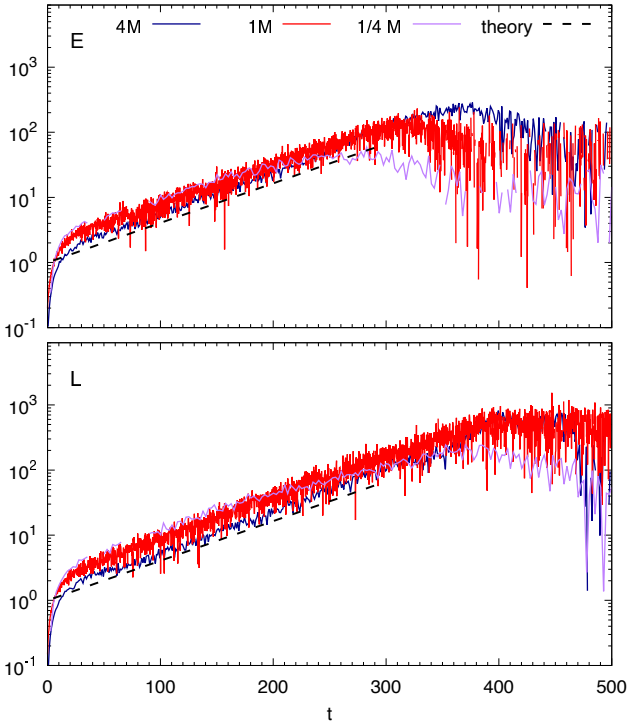}}
  \caption{Normalized diffusion coefficients (Eqs.\,\ref{eq:DC}) for energy (upper) and angular momentum (lower) in three runs with 1/4, 1 and 4 million particles. The dashed lines show the predicted theoretical slope from Fig.\,\ref{fig:a2}.}
  \label{fig:del}
\end{figure}

Fig.\,\ref{fig:del} presents the normalized diffusion coefficients:
\begin{equation}
    D[E] \equiv \frac{t_{\rm rel}}{W^2_{\rm C}} \frac{\d }{\d t}\sigma^2_t[E]\,,\quad D[L] \equiv \frac{t_{\rm rel}}{L^2_{\rm circ}} \frac{\d }{\d t}\sigma^2_t[L]\,.
    \label{eq:DC}
\end{equation}
By construction, it should give constant values $\zeta$ and $\eta$ for two-body or resonant relaxation. In reality, we see an exponential growth of the coefficients with a rate close to that predicted for the quadrupole harmonics (see Fig.\,\ref{fig:a2}). Curves for different $N$ nearly overlap each other, and thus the spreads increment as:
\begin{equation}
    \d \sigma^2_t \propto N^{-1}\,\exp (\gamma t)\,\d t\,.
\end{equation}
From this dependence, we conclude that the accelerated diffusion is due to the instability. The expected contribution of two-body relaxation is less over time and thus cannot be identified. In turn, the resonant relaxation should in principle be seen at the beginning of the simulations up to $t\sim 100$, but the $D[L]$ curves obviously show no sign of this type of relaxation.

\section{Conclusions and final remarks}
\label{sec:summary}

The gravitational loss-cone instability (gLCI) predicted earlier theoretically, is now revealed for the first time in numerical simulations. In the limit of slow precessing orbital motion, allowed {\it for all stars} of a system in dominating Keplerian and harmonic potentials only, it proved possible to connect the sign of the orbital precession rate with the derivative of the DF with respect to the angular momentum. Namely, gLCI is possible if $\p F/\p L\cdot \Omega_{\rm pr}<0$. For the dominating Keplerian potential, the precession is always retrograde, i.e. $\Omega_{\rm pr}<0$. So, independently of the cluster DF, gLCI requires $\p F/\p L > 0$ at small $L$. In the dominating harmonic potential, the precession could be both prograde and retrograde. The latter occurs only if the DF grows sufficiently fast at small angular momenta ($n>0.4$ for the power law DF of Eq.\,\ref{eq:df}). Thus for both types of dominating potentials, the instability requires a deficit of particles at small $L$, i.e. the loss cone must be present. This instability has a well-known counterpart in plasma physics called loss-cone instability~\citep{1965PhFl....8..547R}.

For the first simulations, we choose a toy model in the harmonic potential, since it avoids { complications of hard encounters} with the massive central body. The main feature here is the lengthy duration of simulations, which is 5,000\,--\,10,000 typical dynamical times. Our runs consist of up to 4\,M (million particles), although satisfying results could be obtained already with 0.25\,M. We confirm the value of the growth rate found previously using linear perturbation theory~\citep{2010AstL...36..175P}, but we also note that the shape of the cluster does not change due to early instability saturation. { In addition, we found that other spherical harmonics have growth rates similar to the quadrupole harmonic. } A further study shows the presence of abnormally high diffusion of particles in the $(E,L)$ phase space, obviously connected to the instability. The efficiency of two-body relaxation is smaller over the whole period of simulation. { The resonant relaxation, which is possible in principle for these kinds of stellar systems, seems to be suppressed.}

Although in theory, it is hard to link the instability to the deficit of particles at the lower angular momentum end an arbitrary halo, we followed the instability by gradually decreasing the impact of the halo. It occurs that all features, including the instability growth rate and it's attenuation due to accelerated diffusion, persist for the moderate and even vanishing halos.

In our simulations, the stars didn't scatter out, contrary to the particles in plasma traps where they continuously escape in the direction parallel to the magnetic field. Therefore, the loss cone filling in the stellar cluster leads to cease the instability.

Our next step is to simulate the gLCI with a toy model in the Keplerian dominating potential.

\section*{Acknowledgements}

This work was supported by the Deutsche Forschungsgemeinschaft (DFG, German Research Foundation) -- Project-ID 138713538 -- SFB 881 (``The Milky Way System'', subproject A06), by the Volkswagen Foundation under the Trilateral Partnerships grants No. 90411, and the Basic Research program II.16 (Ilia Shukhman). Peter Berczik acknowledges support by the Chinese Academy of Sciences through the Silk Road Project at NAOC, through the ``Qianren'' special foreign experts program, and the President's International Fellowship for Visiting Scientists program of CAS, the National Science Foundation of China under grant No. 11673032 and also the Strategic Priority Research Program (Pilot B) ``Multi-wavelength gravitational wave universe'' of the Chinese Academy of Sciences (No. XDB23040100). The special GPU accelerated supercomputer Laohu at NAOC has been used and we thank the Center of Information and Computing of NAOC for support. Peter Berczik also acknowledges the special support by the NASU under the Main Astronomical Observatory GRID/GPU computing cluster project. This work benefited from support by the International Space Science Institute, Bern, Switzerland,  through its International Team programme ref. no. 393 ``The Evolution of Rich Stellar Populations \& BH Binaries'' (2017-18).

\appendix
\section{Spherical harmonics in N-body simulations}
\label{app1}

We start with a smooth density distribution
\begin{equation}
\rho(r,\theta,\varphi)=\sum\limits_{l=0}^{\infty} \rho_l(r,\theta,\varphi)\,,
\label{eq:rho}
\end{equation}
where
\begin{equation}
\rho_l(r,\theta,\varphi)=\sum\limits_{m=-l}^l C_l^m(r)\, Y_l^m(\theta,\phi)\,,
\end{equation}
and $Y_l^m(\theta,\phi)$ are fully normalised spherical harmonics:
\begin{multline}
\int d\Omega\, Y_l^m(\theta,\varphi)[Y_{l'}^{m'}(\theta,\varphi)]^* \equiv\\ 
\equiv  \int\limits_0^{\pi} \sin\theta\, d\theta\int\limits_0^{2\pi} d\varphi \,Y_l^m(\theta,\varphi) [Y_{l'}^{m'}(\theta,\varphi)]^*= \delta_{l l'}\,\delta_{m m'}
\end{multline}
($[...]^*$ denotes the complex conjugate). This orthogonality gives the coefficients of the expansion: 
\begin{equation}
C_l^m(r)= \int d\Omega\, \rho_l(r,\theta,\varphi)\, [Y_l^m(\theta,\varphi)]^*,\ \ |m|\le l \,.
\end{equation}
In models of clusters with a DF depending on $E$ and $L$ only, the time dependence (i.e. eigenfrequencies of oscillations) is independent of $m$~\citep[e.g.,][]{1984sv...bookQ....F}.

Now we consider a particle distribution in N-body simulations 
\begin{equation}
\rho(r,\theta,\phi)= \sum\limits_{i=1}^N \mu_i \frac{\delta(r-r_i)}{r^2}\,\frac{\delta(\theta-\theta_i)}
{\sin\theta}\,\delta(\varphi-\varphi_i)
\end{equation}
(here $N$ is the total number of particles; $\mu_i$, $r_i$, $\theta_i$ and $\varphi_i$ are mass and spherical coordinates of particle $i$) and introduce a {\it global} characteristic of each harmonic component as follows:
\begin{equation}
A_l^m = \frac{1}{M_*} \int \d r  \, r^2  C_l^m(r)\, = \frac{1}{M_*}\sum\limits_{i=1}^N \mu_i [Y_l^m(\theta_i,\varphi_i)]^*\,.
\end{equation}
The strength of the spherical harmonic $l$ can be described by coefficients $A_l$, where:
\begin{equation}
A_l^2 \equiv \sum_{m=-l}^l |A_l^m|^2\,. 
\label{eq:str}
\end{equation}
Using the addition theorem for spherical harmonics~\citep[e.g.][]{BatTop}:
\begin{equation}
P_l(\cos\Theta) = \frac{4\pi}{2l+1} \sum_{m=-l}^l Y_l^{m*} (\theta', \varphi') Y_l^{m} (\theta, \varphi)\,,
\end{equation}
($P_l(x)$ are the Legendre polynomials, $\Theta$ is an angle between directions $(\theta, \varphi)$ and $(\theta', \varphi')$), one can show the independence of $A_l$ (Eq.\,\ref{eq:str}) of the orientation of the coordinate frame:
\begin{align}
	N^{2}A_l^2 ={}&  N^{2} \sum_{m=-l}^l |A_l^m|^2 = \notag \\ 
  ={}& \sum_{m=-l}^l \sum\limits_{i=1}^N \mu_i Y_l^{m} (\theta_i, \varphi_i) 
	             \sum\limits_{j=1}^N \mu_j Y_l^{m*}(\theta_j, \varphi_j) = \notag \\
  ={}& \sum\limits_{i=1}^N \sum\limits_{j=1}^N \mu_i \mu_j 
                     \sum_{m=-l}^l Y_l^{m}(\theta_i, \varphi_i) Y_l^{m*}(\theta_j, \varphi_j) = \notag \\
  ={}& \frac{2l+1}{4\pi} \sum\limits_{i=1}^N \sum\limits_{j=1}^N \mu_i \mu_j P_l(\cos\Theta_{ij}) = \notag \\
  ={}& \frac{2l+1}{4\pi} \left[ \sum\limits_{i=1}^N \mu^2_i  + 
                                \sum\limits_{i\ne j} \mu_i \mu_j P_l(\cos\Theta_{ij}) \right]\,
\end{align}
($\Theta_{ij}$ denotes an angle between particle $i$ and $j$). The last expression is clearly independent of the orientation of the frame. If masses of the particles are equal, $\mu_i = M_*/N$, then
\begin{equation}
A_l = \frac1N \left[ \sum_{m=-l}^l \left| \sum\limits_{i=1}^N Y_l^{m} (\theta_i, \varphi_i)\right|^2 \right]^{1/2}\,.
\end{equation}
This expression is used to evaluate the strength of spherical harmonics in our N-body simulations.


\begin{thebibliography}{}

\bibitem[\protect\citeauthoryear{Antonov}{1973}]{A73} Antonov V. A.,
1973, in Omarov E. G., ed., Dynamics of Galaxies and Star
Clusters. Alma Ata, p. 139 (in Russian)
  [trasnslated  in  1987,
  Structure and Dynamics of Elliptical Galaxies, Ed. by T. de Zeeuw,
Proc. IAU Symp., No. 127 (Reidel, Dordrecht), p. 549]

\bibitem[\protect\citeauthoryear{{Batygin} \& {Toptygin}}{{Batygin} \& {Toptygin}}{1978}]{BatTop}
Batygin V.V., and Toptygin I.N., 1978, Problems in Electrodynamics, 2nd ed., Academic Press Inc, London

\bibitem[\protect\citeauthoryear{{Binney} \& {Tremaine}}{{Binney} \&
  {Tremaine}}{2008}]{BT08}
{Binney} J.,  {Tremaine} S.,  2008, {Galactic Dynamics: Second Edition}.
Princeton University Press

\bibitem[\protect\citeauthoryear{{Fridman} \& {Polyachenko}}{{Fridman} \&
  {Polyachenko}}{1984}]{1984sv...bookQ....F}
{Fridman} A.~M.,  {Polyachenko} V.~L.,  1984, {Physics of gravitating systems.
  I - Equilibrium and stability}.
Springer, New York

\bibitem[\protect\citeauthoryear{{Khoperskov}, {Di Matteo}, {Gerhard}, {Katz},
  {Haywood}, {Combes}, {Berczik} \& {Gomez}}{{Khoperskov}
  et~al.}{2019}]{2019A&A...622L...6K}
{Khoperskov} S.,  {Di Matteo} P.,  {Gerhard} O.,  {Katz} D.,  {Haywood} M.,
  {Combes} F.,  {Berczik} P.,    {Gomez} A.,  2019, \aap, 622, L6

\bibitem[\protect\citeauthoryear{{Palmer}}{{Palmer}}{1994}]{1994ASSL..185.....P}
{Palmer} P.~L.,  1994, {Stability of collisionless stellar systems: mechanisms
  for the dynamical structure of galaxies}.
Vol.~185, Kluwer, Dordrecht

\bibitem[\protect\citeauthoryear{{Polyachenko}, {Berczik} \&
  {Just}}{{Polyachenko} et~al.}{2016}]{PBJ16}
{Polyachenko} E.~V.,  {Berczik} P.,    {Just} A.,  2016, \mnras, 462, 3727

\bibitem[\protect\citeauthoryear{{Polyachenko}, {Polyachenko} \&
  {Shukhman}}{{Polyachenko} et~al.}{2007}]{2007MNRAS.379..573P}
{Polyachenko} E.~V.,  {Polyachenko} V.~L.,    {Shukhman} I.~G.,  2007, \mnras,
  379, 573

\bibitem[\protect\citeauthoryear{{Polyachenko} \& {Shukhman}}{{Polyachenko} \&
  {Shukhman}}{2015}]{2015MNRAS.451..601P}
{Polyachenko} E.~V.,  {Shukhman} I.~G.,  2015, \mnras, 451, 601

\bibitem[\protect\citeauthoryear{{Polyachenko}}{{Polyachenko}}{1991}]{1991SvAL...17..371P}
{Polyachenko} V.~L.,  1991, Soviet Astronomy Letters, 17, 371

\bibitem[\protect\citeauthoryear{{Polyachenko}, {Polyachenko} \&
  {Shukhman}}{{Polyachenko} et~al.}{2010}]{2010AstL...36..175P}
{Polyachenko} V.~L.,  {Polyachenko} E.~V.,    {Shukhman} I.~G.,  2010,
  Astronomy Letters, 36, 175

\bibitem[\protect\citeauthoryear{{Polyachenko} \& {Shukhman}}{{Polyachenko} \&
  {Shukhman}}{1972}]{PS1972}
{Polyachenko} V.~L.,  {Shukhman} I.~G.,  1972, Preprint SibIZMIR 1-2-72,
  Stability of gravitating systems with quadratic potential. Parts I and II,
  Irkutsk (in Russian)

\bibitem[\protect\citeauthoryear{{Polyachenko} \& {Shukhman}}{{Polyachenko} \&
  {Shukhman}}{1981}]{PS1981SvA}
{Polyachenko} V.~L.,  {Shukhman} I.~G.,  1981, \sovast, 25, 533

\bibitem[\protect\citeauthoryear{{Rauch} \& {Tremaine}}{{Rauch} \&
  {Tremaine}}{1996}]{1996JRASC..90..334R}
{Rauch} K.~P.,  {Tremaine} S.,  1996, \jrasc, 90, 334

\bibitem[\protect\citeauthoryear{{Read}, {Goerdt}, {Moore}, {Pontzen}, {Stadel}
  \& {Lake}}{{Read} et~al.}{2006}]{2006MNRAS.373.1451R}
{Read} J.~I.,  {Goerdt} T.,  {Moore} B.,  {Pontzen} A.~P.,  {Stadel} J.,
  {Lake} G.,  2006, \mnras, 373, 1451

\bibitem[\protect\citeauthoryear{{Rosenbluth} \& {Post}}{{Rosenbluth} \&
  {Post}}{1965}]{1965PhFl....8..547R}
{Rosenbluth} M.~N.,  {Post} R.~F.,  1965, Physics of Fluids, 8, 547

\bibitem[\protect\citeauthoryear{{Sellwood}}{{Sellwood}}{2015}]{2015MNRAS.453.2919S}
{Sellwood} J.~A.,  2015, \mnras, 453, 2919

\bibitem[\protect\citeauthoryear{{Tremaine}}{{Tremaine}}{2005}]{2005ApJ...625..143T}
{Tremaine} S.,  2005, \apj, 625, 143

\bibitem[\protect\citeauthoryear{{Zinchenko}, {Berczik}, {Grebel}, {Pilyugin}
  \& {Just}}{{Zinchenko} et~al.}{2015}]{2015ApJ...806..267Z}
{Zinchenko} I.~A.,  {Berczik} P.,  {Grebel} E.~K.,  {Pilyugin} L.~S.,    {Just}
  A.,  2015, \apj, 806, 267


\end{thebibliography}
\end{document}